\documentclass[preprint,showpacs,preprintnumbers,amsmath,amssymb,nofootinbib]{revtex4}

\usepackage{graphicx}
\usepackage{epsfig}
\usepackage{bm}
\usepackage{amsfonts}
\usepackage{subfigure}
\usepackage{multirow}
\usepackage{color}
\usepackage{float}

\begin{document}



\title{\textbf{Higher order mimetic gravity after GW170817}}



\author{Kimya Sharafati\footnote{kimiasharafati@gmail.com}, Soma Heydari\footnote{s.heydari@uok.ac.ir} and Kayoomars Karami\footnote{kkarami@uok.ac.ir}}
\address{\small{Department of Physics, University of Kurdistan, Pasdaran Street, P.O. Box 66177-15175, Sanandaj, Iran}}

\date{\today}

\begin{abstract}
On the 17th of August 2017,  the thriving discovery of  gravitational waves event GW170817 and its optical counterpart GRB170817A, owing to coalescing of two neutron starts divulged a  very small amount of difference around  ${\cal O}(10^{-16})$ between traveling speed of light and the velocity of gravitational waves ($C_{T}$). This small deviation can be used as a strong constraint on modified gravity models. We concentrate on the Higher-Order expansion of Mimetic Gravity (HOMimG) model to specify the parametric space of three parameters of our model ($a$, $b$, and $c$ ) utilizing the observational constraint from GW170817-GRB170817A on $C_{T}$, besides two theoretical constraints on $C_{T}^{2}$ and $C_{s}^{2}$ due to assurance of the stability of the model and  subluminal promulgation of the scalar and tensor perturbations.  Thereafter, we increase the accuracy of the parametric space with the aid of imposing further limitation of $\gamma$ parameter (related to the age of the Universe). In pursuance of determining the other parameter of the model ($\lambda$), the potential of the model is specified, and another observational bound related to the Equation of State (EoS) parameter of dark energy is taken into account. In consequence, we attain a viable HOMimG model confined to numbers of observational and theoretical constraints. At the end, regarding the concluded numerical ranges for the model parameters, and cogitating two different potential  (quadratic and quartic potentials) to specify $\lambda$ parameter, we illustrate that the values of the model parameters are independent of the form of potential.
\end{abstract}



\maketitle

\newpage
\section{Introduction}
Almost a hundred years after the birth of General Relativity (GR) in September 2015, discovery of gravitational waves (GWs) emitted from the collision between two black holes by LIGO-VIRGO collaboration, was a preface to construct the GWs astronomy~\cite{Abbott:2016-a} as a new subdivision of astronomy.
Thereafter, some other analogous events from coalescing binary black holes were detected \cite{Abbott:2016b,Abbott:2017a,Abbott:2017b,Abbott:2017c,Abbott:2019}. On the 17th August 2017, a disparate detection caused by the merger of two neutron stars namely GW170817 event was accomplished by
the LIGO-VIRGO  collaboration  ~\cite{Abbott:2017d}, and its optical counterpart viz GRB170817A by a time difference of $\delta t=(1.734 \pm 0.054)\,s$ was discovered by Fermi Gamma-Ray Space Telescope \cite{Goldstein:2017}. Most of this time difference due to astrophysical donations, is related to  crumple of the ponderous neutron star generated through the coalition. The amount of these astrophysical donations was assessed as about $\approx 10\,s$ in~\cite{Abbott:2017e}.
Thenceforth, the researchers have taken into account the comparison between  velocity of light and the traveling speed of GWs in their investigations.

Furthermore, the observational data of the Ia Super-Novae \cite{Riess:1998,Perlmutter:1999}, large scale structure \cite{Tegmark:2004}, Cosmic Microwave Background (CMB) anisotropies \cite{Spergel:2003}, beside week lensing \cite{Jain:2003} have corroborated that we have an expanding accelerating Universe. So as to vindicate this observational outcome, two assortments of  substitute theories for GR were existed. The first assortment, consists in  definition of strange substance with negative pressure and anti-gravity characteristics, namely dark energy theories \cite{Padmanabhan:2003,Copeland:2006,Li:2011}. The second assortment, is based on amending the
geometric term of the GR action, to wit modified gravity such as scalar-tensor theories \cite{Horndeski:1974,Clifton:2012}, $f({\cal G})$ \cite{Nojiri:2005,De Felice:2010}, $f(T)$ \cite{Cai:2016}, $f(R)$ \cite{Sotiriou:2010}, and etc.\\
The mysterious  nature of dark energy~\cite{Frieman:2008sn} and dark matter~\cite{Garrett:2010hd}, beside the giant fine-tuning problem of cosmological constant ($\Lambda$) in the standard model of cosmology ($\Lambda$CDM) \cite{Weinberg:1988cp},  inspire the researchers  to achieve alternative scenarios for GR, so as to explain the  expansion history of the present universe without dark matter, constituting solely of standard baryonic matter and radiation ~\cite{Nojiri:2006ri,Nojiri:2010wj,Clifton:2012,Capozziello:2011et,Nojiri:2017ncd}. Many attempts have been made to overpower these problems by considering a dynamic source for the present expanding accelerating universe  ~\cite{Copeland:2006,Bamba:2012cp}.

After the prosperous detection of GWs (GW170817-GRB170817A events), the multifarious modified gravity and dark energy theories in  view of  LIGO-VIRGO  data  have been inquired by researchers~\cite{Ezquiaga,Amendola,Baker,Sakstein,Cai:2018,Akrami:2018,Langlois:2018,Kase:2018,Crisostomi:2018,Oost:2018,
Gong:2018,skarimi,Boran}. It is well understood that, exact determination of the traveling speed  of GWs ($C_T$) surprisingly could place strict constrains on modified gravity theories, because their speed has an infinitesimal deviation from the speed of light ($C_T= 1$). Hence, modified gravity models which predict $C_T\neq 1$ for the traveling speed of GWs should be seriously re-examined.


It is known that, several ways to modify GR are suggested by scalar-tensor theories with the second order equations of motion and coupling between scalar fields and gravity, so-called Horndeski's theory \cite{Horndeski:1974}. The Horndeski gravity supplies a generic action evading the Ostrogradski instabilities \cite{Ostro1, Ostro2,Ostro3}. The compatibility of the Horndeski and beyond Horndeski theories with the LIGO-VIRGO data  has been investigated in \cite{Ezquiaga}.

Mimetic Gravity (MimG) as a special specimen of higher order scalar-tensor theories  suggested by Chamseddine and Mukhanov  ~\cite{Chamseddine:2013kea} (for important earlier works, see \cite{Lim:2010yk,Capozziello:2010,Gao:2010gj,Zumalacarregui:2013pma}), is related to GR with a mimetic scalar field $\phi$, and a non-invertible disformal transformation of the metric~\cite{Deruelle:2014zza,Domenech:2015tca,Achour:2016rkg}.
Latterly in ~\cite{Ezquiaga},  MimG model was considered as a specific case of a Degenerate Higher Order Scalar-Tensor (DHOST) theory~\cite{Zumalacarregui:2013pma,Ostro3} (for related discussions on the relation between MimG and DHOST theories, see \cite{Achour:2016rkg,Langlois:2018jdg}) attained from conformal transformation of Horndeski theory with $C_T= 1$.

Here,  we concentrate on a particulary interesting model of the Higher Order expansion of MimG (HOMimG), proposed by  Cognola \textit{et al.} ~\cite{Cognola:2016gjy}, and try to specify the range of model parameters. The motivation of choosing this alternative is the attractive feature of the model which is exposed in the low energy limit of the Ho$\check{r}$ava-Lifshitz gravity. Note that Casalino \textit{et al.} ~\cite{Casalino:2019} by using the observational constraint from GW170817/GRB170817A on the GW speed and imposing the absence of both ghost instabilities and superluminal propagation
of scalar and tensor perturbations estimated the upper limits on the three free parameters of HOMimG model.

In the present work, firstly like Casalino \textit{et al.} ~\cite{Casalino:2019} we impose the observational bound on the speed of GWs deduced from LIGO-VIRGO events GW170817/GRB170817A, as well as two theoretical restrictions to avoid of instability and superluminality to find constraints on the three parameters of HOMimG model. Then, using the additional observational constraint of the age of the universe, we increase the accuracy of parametric space of the model. After all, we apply another observational constraint of the equation of state parameter of dark energy, to find the limitation on the fourth parameter of the model. Consequently, we will be able to  constrain viable extended HOMimG model in the light of observational data and theoretical considerations.

The rest of this article is organized as follows: In Sec.~\ref{sec:II}, we concisely study the extended HOMimG model, thereto  background  and late-time cosmological evolution. In the following the stability conditions  and parametric space of the model according to theoretical and observational constraints are determined in Sec.~\ref{sec:IV}. Finally, we epitomize our conclusions in Sec.~\ref{sec:V}.

\section{Higher Order expansion of Mimetic gravity}\label{sec:II}
 Mimetic gravity action can be attained, by inaugurating from the Einstein-Hilbert action and parameterizing the physical metric $g_{\mu \nu}$ with respect to an auxiliary metric $\tilde{g}_{\mu \nu}$ and the mimetic field $\phi$ as
\begin{eqnarray}
g_{\mu \nu} = -\tilde{g}_{\mu \nu}\tilde{g}^{\alpha\beta}\partial_{\alpha}\phi\partial_{\beta}\phi \, .
\label{mimetictransformation}
\end{eqnarray}
Note that to achieve consistency,  the following mimetic proviso should be gratified
\begin{eqnarray}
g^{\mu \nu}\partial_{\mu}\phi\partial_{\nu}\phi = -1 \ .
\label{mimeticconstraint}
\end{eqnarray}By taking variation of action  with regard to $g_{\mu\nu}$, considering its association with $\tilde{g}_{\mu \nu}$ and $\phi$ by way of Eq. (\ref{mimetictransformation}), the field equations can be obtained with an additional term that acts like as pressureless fluid and it can be considered as mimetic cold dark matter \cite{Chamseddine:2013kea}. Furthermore, It is proven that mimetic gravity could explain galactic rotation curves in order to fully match the phenomenology of dark matter \cite{Myrzakulov,Vagnozzi}.
Moreover, it has been established that, a very intelligible expansion of original MimG
with a potential $V(\phi)$ for the mimetic field, not only can imitate dark energy at late time, but also can  supply  early-time inflationary epoch  ~\cite{Chamseddine}.

It is worth mentioning that, the  mimetic models have trouble with  some sharp instabilities on  small scales. Therefore the proper "completion"  is required to eliminate these instabilities and accommodate the theory mathematically \cite{Capela:2015,Mirzagholi:2015,Leon:2015}.

In the earliest MimG model, the sound speed $C_s$ (traveling speed of the scalar perturbations) is zero~\cite{Chamseddine}, which gives rise to complexity of defining quantum fluctuations of the mimetic field. Latterly it is explained that, this problem remains in the mimetic Hordensky gravity as well~\cite{Arroja:2015yvd}. In this regard, recently  an explicit mimetic Hordensky model has been studied in \cite{Cognola:2016gjy}, to obtain a Higher-Order Mimetic Gravity model
with $C_s \neq 0$, through definitely  fracturing the structure of the original Horndeski model.

In the present work, we contemplate the HOMimG model for two reasons. Firstly, it emerges from the low energy limit of  Ho$\check{r}$ava-Lifshitz gravity that is a good candidate for quantum gravity, and secondly, eliminated instabilities in this model through the obtained nonzero sound speed for scalar perturbation ($C_s \neq 0$)~\cite{Cognola:2016gjy}.  The action of the HOMimG model ~\cite{Cognola:2016gjy}, is as follows (see also~\cite{Rinaldi:2016oqp,Diez-Tejedor:2018fue,Casalino:2019,Sebastiani} for related studies)
\begin{eqnarray}
S = \frac{1}{2}\int d ^4x\sqrt{-g} [ R(1 + a g^{\mu \nu}\nabla_{\mu}\phi\nabla_{\nu}\phi) - \frac{c}{2}(\Box \phi)^2 \nonumber \\
+ \frac{b}{2} (\nabla_{\mu}\nabla_{\nu}\phi)^2 - \frac{\lambda}{2} \left ( g ^{\mu \nu}\nabla _{\mu}\phi\nabla _{\nu}\phi + 1 \right )  - V (\phi) ] \,.
\label{cmsvz}
\end{eqnarray}
wherein $g$ is the determinant of the metric tensor $g_{\mu\nu}$, $R$ is the Ricci scalar, $\phi$ is the mimetic field, $\lambda$ is the Lagrangian multiplier (acquainted to impose the mimetic proviso~(\ref{mimeticconstraint}) on the mimetic field), $V(\phi)$ is the potential of the mimetic field, and  $a$, $b$, and $c$ are constant parameters. It is worth noting that, for the case $b=c=4a$ the mimetic Horndeski model is recovered \cite{Horndeski:1974}.
\subsection{background equations}
\label{background}
 We pursue the calculations of ~\cite{Casalino:2018tcd}, and study the background evolution with the
flat Friedmann-Robertson-Walker (FRW) metric and generally positive signature $g_{{\mu}{\nu}}={\rm diag}\big(-1, A^{2}(t), A^{2}(t), A^{2}(t)\big)$,
where $A(t)$ is the scale factor and $t$ is the cosmic time.\\
The mimetic restriction (\ref{mimeticconstraint}) on this background results in the equality between  mimetic field and cosmic time as follows
\begin{eqnarray}\label{mim_phi}
\phi=t\,.
\end{eqnarray}
Thereafter, by taking variation of the action (\ref{cmsvz}) with respect to $g_{\mu\nu}$, the Friedmann equations are obtained as follows 
\begin{align}
&(-6c+24a)\dot H+(36a+9c+12-9b)H^2-2V-2\lambda -2\rho_{\rm m}=0\,,\label{ttcomp}\\
&3H^2+2\dot H={2V-2P_{\rm m}\over 4-4a-b+3c}\,,\label{final}
\end{align}
wherein $H\equiv \dot{A}/A $ is the Hubble parameter,  a dot signifies derivative against the cosmic time $t$, and moreover  $P_{\rm m}$ and $\rho_{\rm m}$ correspond to the  pressure and energy density of the compound of baryonic matter and radiation.

Note that with the following definitions
\begin{align}
\rho_\text{df}\equiv\lambda+V+(3 b - 24 a) H^2 + (3 c - 12 a)\dot H\label{rhoP}\,,
\end{align}
 \begin{align}
  P_\text{df}\equiv-V\,,\label{PP}
  \end{align}
one can recast the Friedmann equations (\ref{ttcomp})-(\ref{final}) in the following usual forms
\begin{eqnarray}\label{friedmann}
6H^2 &=&\frac{4}{4-b + 3c - 4a}\, \left(\rho_\text{df} + \rho_\text{m}\right) \,,\\
-4\dot H - 6H^2&=&\frac{4}{4-b + 3c - 4a} \left(P_\text{df} + P_\text{m}\right)\,,
\end{eqnarray}
where $\rho_\text{df}$ and $P_\text{df}$ are effective energy density and pressure of dark fluid (containing dark energy and dark matter), respectively.

The sound speed ($C_s$) and gravitational wave speed ($C_T$) associated with the propagated scalar and tensor perturbations are deduced via appropriate perturbation of FRW background as follows \cite{Casalino:2018tcd}
\begin{eqnarray}
\label{cs}
C_s^2&=&\frac{2(b-c)(a-1)}{(2a-b-2)(4-4a-b+3c)}\,, \\
C_{T}^{2}&=&\frac{2(1-a)}{2(1-a)+b}\,.
\label{ct}
\end{eqnarray}
It is clear from Eq. (\ref{ct}) that, in order to obtain $C_T\neq1$ the parameter $b$ must be nonzero  ($b\neq0$). Note that for $a=b=c=0$ we have $C_s^2=0$ and $C_T^2=1$, hence the primary MimG model in admirable consistency with the GW170817-GRB170817A observations is recovered.
In \cite{Casalino:2018tcd} the quadratic action for scalar and tensor perturbations have been calculated and liberated  theory from ghosts problem has been obtained, provided that $c>0$ and $a<1$. Furthermore the steadiness guarantee inflicts  theoretical  requirements as $0 \leq C_T^2 \leq 1$ and $0 \leq C_s^2 \leq 1$, in which the upper bound implements avoidance of superluminal dissemination on scalar and tensor perturbations \cite{Hirano:2017zox,Zheng:2017qfs,Cai:2017dyi,Takahashi:2017pje,Gorji:2017cai}.
The late approximate simultaneous observatory of GW170817~\cite{Abbott:2017d} and its electromagnetic counterpart GRB170817A~\cite{Abbott:2017e}, has appointed strict restrictions on the speed of GWs in comparison with the velocity  of light.  Thus we contemplate the  following LIGO-VIRGO constraint on $C_{T}$~\cite{Abbott:2017e}, to impose restriction  on the parameters of the HOMimG model.
\begin{eqnarray}
1-3\times10^{-15} \leq C_T\leq 1+7\times10^{-16}.
\label{LIGO-VIRGO}
\end{eqnarray}
In the following, we analyze some conditions to have appropriate HOMimG model applying background and late-time cosmological equation.

It is clear from the action~(\ref{cmsvz}) that, the multiplied term to the Ricci scalar ($1+ag^{\mu \nu}\nabla_{\mu}\phi\nabla_{\nu}\phi = 1-a$) can be considered as effective Newton constant. Ergo $a<1$ must be confirmed, in order to certify the positivity of the Newton constant, and as likewise  mentioned  previously,  $a<1$ is required  to eliminate the ghosts problem from the model.
Additionally, it is known that $\vert a \vert \gg{\cal O}(1)$ can be intricate in the theory of ultraviolet complete gravity, so with this in mind and regarding
the perturbation discussions, the expected condition for parameter $a$ is  $\vert a \vert \lesssim {\cal O}(1)$.
Ultimately, according to above explanations, and for directness of calculations we consider the range of  $(-1,1)$ for the parameter $a$.

Apropos of $b$, it is clear from Eq. (\ref{ct}) that, to satisfy the observational restriction of  LIGO-VIRGO collaboration (\ref{LIGO-VIRGO}) by  GWs speed, the parameter $b$ must be $\ll {\cal O}(1)$. Furthermore, one can see from  Eq. (\ref{ct}) that, $C_{T}^{2}>1$ would be obtained if $b<0$, in which the subluminality necessity for $C_{T}$ is contravened . Thus, according to the above discussions, the parameter $b$ is constrained to be in the range of $(0,1)$.

The next part of this analysis is related to the parameter $c$. As it was mentioned in the preceding descriptions, from the obtained sound speed equation~(\ref{cs}), the condition $c>0$ can give rise to emancipate the model from ghosts problem. Thereupon, regarding the subluminality requisite for the sound speed, the parameter $c$ is confined to be in the range of $(0,1)$.

\subsection{Late-time cosmological evolution}

\label{late_time}
By substituting $\dot H=-H^{2}(1+q)$ in Eq. (\ref{rhoP}), and using Eq. (\ref{PP}), the effective Equation of State (EoS) parameter of the dark fluid is obtained as follows
\begin{equation}
\omega_\text{df}\equiv\frac{P_\text{df}}{\rho_\text{df}} = \frac{-V}{V+\lambda - 24 a H^2 - (3 c - 12 a)(1+q) H^{2}} \,.\label{eos}
\end{equation}
Note that $q$ is the decelerating cosmographic parameter with the present value of $q=-1.2037\pm 0.175$ at 68\% CL \cite{Capozziello:2019}.
The latest observations of CMB data associated with  polarization and temperature anisotropies and cross-correlations of them, beside the geometric evaluations of the luminosity distance of Supernova Type-Ia, in addition to the appraisement of Baryon Acoustic Oscillations propose that, EoS parameter of dark energy has a minute digression from the amount of the related parameter to cosmological constant ($\omega=-1$). The observational constraint on the EoS parameter of dark energy is given by  Planck 2018 TT,TE,EE+lowE+lensing+BAO data at  $68\%$ CL \cite{Planck:2018},  as follows
\begin{equation}
-1.04-0.1<\omega_{\rm de} < -1.04+0.1.
\label{omega_df}
\end{equation}
\begin{figure}[H]
\centering
\includegraphics[scale=0.55]{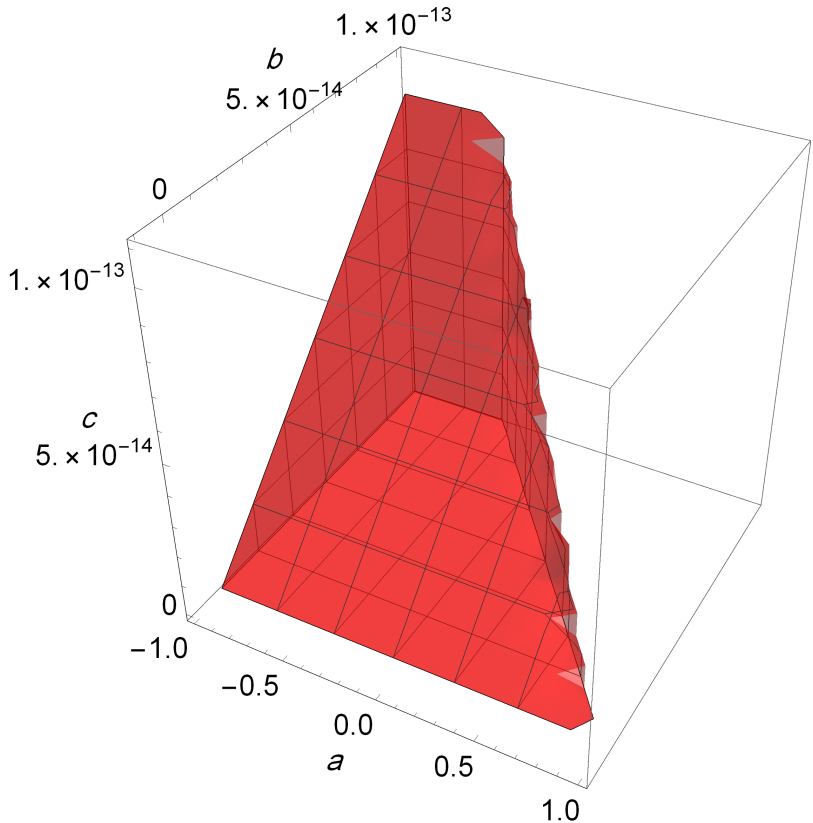}
\vspace{-0.5em}
\caption{Parametric space for $a$, $b$ and $c$, in HOMimG model, by establishing the constraints on $C_s^2$, $C_T^2$, $C_T$.}
\label{fig1}
\end{figure}
\noindent
Another quantity that can be used to impose observational constraints on the model parameters is the age of the universe $t_0$. Hence, we consider the following relation between $t_{0}$ and the parameters $a$, $b$, and $c$ as \cite{Casalino:2018tcd}
\begin{equation}
t_0 = \int_0^1 \, \frac{dA}{A H(A)} = \sqrt{\frac{4-b+3c-4a}{4}}\int_0^1\, \frac{dA}{A H_0 \sqrt{\frac{1}{6}\sum_i \Omega_i(A)}}\,,
\end{equation}
wherein $H_{0}=H(t_{0})$ is the current Hubble parameter. It is clear from the above equation that, for $4a=3c$ and $b=0$ or in the specimen of $a=b=c=0$ (related to GR) the preceding coefficient of the integral will be identical to $1$.
Thus the age of the cosmos can be written as the following form
\begin{eqnarray}
t_0 = \sqrt{\frac{4-b+3c-4a}{4}} ~t_0\Big|_{4a=3c,b=0}\equiv \gamma t_0\Big|_{4a=3c,b=0}\,.
\end{eqnarray}
Using the present value for the age of the universe $t_{0}=13.787\pm0.020$, measured by  Planck 2018 TT,TE,EE+lowE+lensing+BAO data at the $68\%$ CL \cite{Planck:2018}, one can easily obtain the following observational range for the parameter $\gamma$
\begin{figure}[H]
\centering
\includegraphics[scale=0.55]{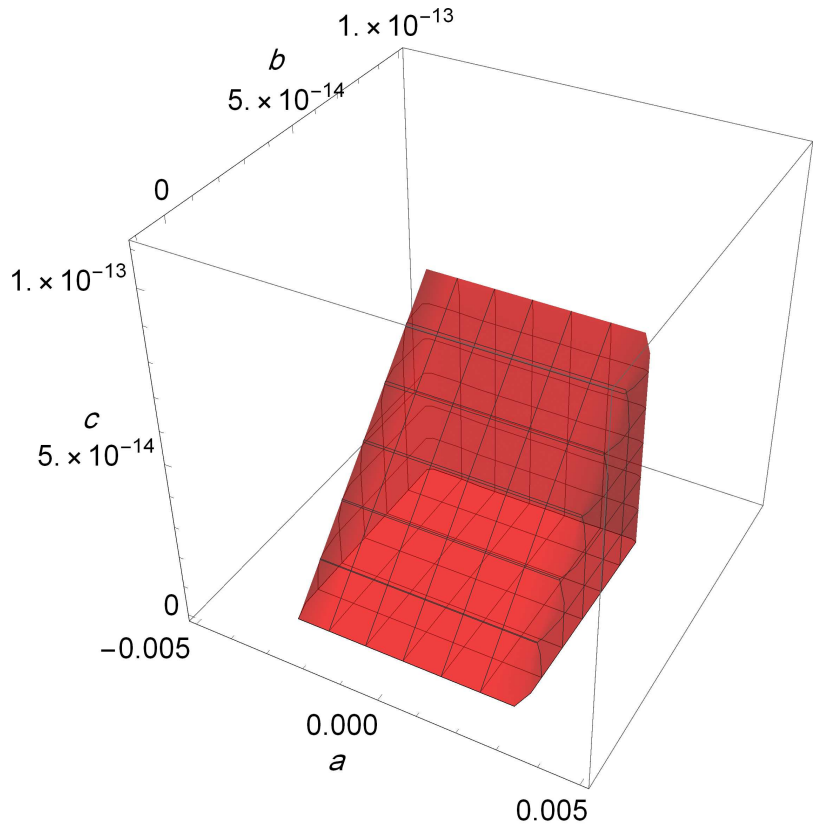}
\vspace{-0.5em}
\caption{Parametric space for $a$, $b$ and $c$, in HOMimG model, by establishing the constraints on $C_s^2$, $C_T^2$, $C_T$ and $\gamma$.}
\label{fig2}
\end{figure}
\noindent
\begin{equation}\label{gammarange}
 \gamma = 1.000 \pm 0.0015.
\end{equation}
\section{Constraining viable HOMimG model}
\label{sec:IV}
Pursuant to our descriptions in the preceding section,  the action (\ref{cmsvz}) of HOMimG model consists of four parameters denoted by $a$, $b$, $c$, and $\lambda$, which can be restricted by numbers of the theoretical and observational constrains, and  using the bounds on these parameters the parametric space of the model can be specified.

In the first step, applying two theoretical conditions on $C_T^2$ and $C_s^2 $ as ${0}\leq C_T^2 \leq{1}$ and  ${0}\leq C_s^2 \leq{1}$, beside the observational constraint of LIGO-VIRGO collaboration (\ref{LIGO-VIRGO}), we can specify the parametric space of $a$, $b$, and $c$. The schemed result is displayed in Fig. \ref{fig1}, and the related numerical bounds for each parameter are listed in the top box of Table \ref{t1}.
\begin{table}[H]
\caption{The attained numerical bounds on parameters of  HOMimG model by considering two  theoretical constraints on $C_s^2$ and $C_T^2$, beside the observational LIGO-VIRGO collaboration constraint on $C_T$, Eq. (\ref{LIGO-VIRGO}), (top box of Table). Deduced numerical bounds on  parameters of HOMimG model by applying two theoretical constraints on $C_s^2$ and $C_T^2$, beside two observational  constraints consist of the limitation of  LIGO-VIRGO collaboration on $C_T$, Eq. (\ref{LIGO-VIRGO}), and $\gamma$ constraint, Eq. (\ref{gammarange}), (bottom box of Table).}
\centering
\begin{tabular}{cc}
\hline
\hline
\quad Constraint \quad & \quad \quad Parameter \quad \quad \quad \quad\\ \hline
\begin{tabular}[c]{@{}c@{}} \quad \quad
${0}\leq C_T^2 \leq{1}$ \quad \quad
\\ \quad ${0}\leq C_s^2 \leq{1}$ \quad \quad
\\ \quad ${1-3\times 10^{-15}}\leq C_T\leq{1+7\times 10^{-16}}$ \quad \quad
\\
\end{tabular} &
\begin{tabular}[c]{@{}c@{}}
\quad \quad ${-1}\leq a \leq 1$  \quad \quad
\\ \quad ${0}\leq b \leq 6.9\times10^{-14}$ \quad \quad
\\ \quad ${0}\leq c \leq 2.39\times10^{-14}$ \quad \quad
\end{tabular} \\ \hline
\begin{tabular}[c]{@{}c@{}} \quad
${0}\leq C_T^2 \leq{1}$ \quad \quad
\\ \quad ${0}\leq C_s^2 \leq{1}$ \quad \quad
\\ \quad  ${1-3\times 10^{-15}}\leq C_T\leq{1+7\times 10^{-16}}$ \quad \quad
\\ \quad ${1-0.0015}\leq \gamma \leq{1+0.0015}$ \quad \quad
\\
\end{tabular} &
\begin{tabular}[c]{@{}c@{}} \quad
 ${-0.003}\leq a \leq 0.0029$  \quad \quad
\\ \quad ${0}\leq b \leq 1.2\times10^{-14}$ \quad \quad
\\ \quad ${0}\leq c \leq 1.2\times10^{-14}$ \quad
\end{tabular}\\
\hline
\hline
\end{tabular}
\label{t1}
\end{table}
\noindent

In the next step, to increase the accuracy of measuring the parametric space, we contemplate further observational constraint (\ref{gammarange}) on $\gamma$ parameter in addition to the three previous
constraints. Accordingly, in this stage we have two observational and two theoretical constraints (as listed in the bottom box of  Table \ref{t1}) to restrict the parameters of our model. So the new numerical restrictions on the parameters $a$, $b$, and  $c$ are represented in the bottom box of  Table  \ref{t1}, and the corresponding parametric space is schemed in Fig. \ref{fig2}.

Heretofore,  we have successfully determined the parametric space, and numerical ranges  for three parameters ($a$, $b$, and $c$) of the action (\ref{cmsvz}). Likewise, in order to specify the range of  $\lambda$ parameter,  a supplementary  constraint beside the previous ones is necessitated. Hereupon, regarding Eq. (\ref{eos}), we apply the latest observational constraint on the EoS parameter of dark energy (\ref{omega_df}), i.e. $\omega_{\rm df}=\omega_{\rm de}$, and  choose a potential for the model. To this end, by selecting the quadratic potential ($\frac{1}{2}m^{2}\phi^{2}$), and using of two theoretical constraints on $C_s^2$ and $C_T^2$ in addition to three observational constraints on $C_T$ (\ref{LIGO-VIRGO}), $\gamma$ (\ref{gammarange}), and  $\omega_{\rm df}$ (\ref{omega_df}), the ranges of four parameters of our model ($a$, $b$, $c$, and $\lambda$) can be computed. Thenceforth, we replicate the calculation of parameter ranges for the quartic potential ($\frac{1}{4} \lambda_{\phi} \phi^{4}$) as well.
\begin{table}[H]
\caption{The deduced ranges for  parameters of HOMimG model with quadratic ($\frac{1}{2}m^{2}\phi^{2}$) and quartic ($\frac{1}{4} \lambda_{\phi} \phi^{4}$) potentials, by considering two theoretical constraints on $C_s^2$ and $C_T^2$,  beside three observational  constraints consist of the limitation of LIGO-VIRGO collaboration on $C_T$ (\ref{LIGO-VIRGO}), $\gamma$ constraint (\ref{gammarange}), and the EoS parameter of dark energy (\ref{omega_df}).}
\begin{tabular}{ccc}
\hline
\hline
\quad Constraint \quad & \quad $\frac{1}{2}m^{2}\phi^{2}$ \quad & \quad $\frac{1}{4} \lambda_{\phi} \phi^{4} $ \quad \\ \hline
\begin{tabular}[c]{@{}c@{}}
${0}\leq C_T^2 \leq{1}$ \quad
\\ \quad ${0}\leq C_s^2 \leq{1}$ \quad
\\ \quad ${1-3\times 10^{-15}}\leq C_T\leq{1+7\times 10^{-16}}$ \quad
\\ \quad ${1-0.0015}\leq \gamma \leq{1+0.0015}$ \quad
\\ \quad ${-1.04-0.1}\leq \omega_{\rm df}\leq{-1.04+0.1}$ \quad
\\
\end{tabular} &
\begin{tabular}[c]{@{}c@{}}
\quad ${-0.003}\leq a \leq 0.0029$  \quad \\
\quad ${0}\leq b \leq 1.2\times10^{-14}$ \quad\\
\quad ${0}\leq c \leq 1.2\times10^{-14}$ \quad\\
\quad $-0.4789\leq \lambda \leq 0.2505 $ \quad
\end{tabular} &
\begin{tabular}[c]{@{}c@{}}
\quad ${-0.003}\leq a \leq 0.0029$  \quad \\
\quad ${0}\leq b \leq 1.2\times10^{-14}$ \quad\\
\quad ${0}\leq c \leq 1.2\times10^{-14}$ \quad\\
\quad $-0.4789\leq \lambda \leq 0.2505 $ \quad
\end{tabular}\\
\hline
\hline
\end{tabular}
\label{t2}
\end{table}
\noindent
The numerical obtained bounds for each parameters related to the selected  potentials are
represented in Table \ref{t2}. The listed results for two different potentials in Table \ref{t2} indicate that, our HOMimG model is independent of the form of potential.

As previously mentioned, Casalino \textit{et al.} ~\cite{Casalino:2019} by using the symmetric bound (for simplicity) on the fractional deviation of the GW speed from the speed of light, $|\delta C_{T}| < 5 \times 10^{-16} $ \cite{Ezquiaga}, beside  theoretical bounds due to absence of both ghost instabilities and superluminal propagation of scalar and tensor perturbations estimated the upper bounds on the model parameters
as $a < 0.27$, $b < 7.18 \times 10^{-15}$, and $c < 4.68 \times 10^{-15}$. In the present paper, we first like Casalino \textit{et al.} ~\cite{Casalino:2019} use the observational bound on the speed of GWs (nonsymmetric case $1-3\times10^{-15} \leq C_T\leq 1+7\times10^{-16}$) as well as two theoretical considerations to enforce the absence of instability and superluminality in the model. This results the restricted bounds on the three parameters ($a,b,c$) of the model (see the top box of Table \ref{t1}). In the following, we apply another observational constraint from the age of the universe to obtain more accurate parametric space for $a,b,c$ (see the bottom box of Table \ref{t1}). In the next, we use further observational constraint on the EoS parameter of dark energy which enables us to restrict the fourth parameter ($\lambda$) of the model (see Table \ref{t2}).

Now, we propose to specify the dependence of  parameters $m$ and $\lambda_{\phi}$, related to quadratic and quartic potential, with respect to the other parameters of the model ($a$, $b$, $c$, and $\lambda$). So as to attain this objective, we consider the following equation for density parameter of dark fluid comprising of dark energy and dark matter as
\begin{equation}\label{frac_density1}
\Omega_{\rm df}(t) = \frac{1}{3 M_{\rm p}^2}\,\frac{\rho_{\rm df}(t)}{ H(t)^2}\,,
\end{equation}
wherein, the present value of density parameter of dark fluid is measured by Planck 2018 TT,TE,EE+lowE+lensing+BAO as $\Omega_{\rm df}(t_0) = 0.95$ \cite{Planck:2018}, and we take the reduced Planck mass equal to one $(M_{\rm p}=1)$. Subsequently, considering Eq. (\ref{rhoP}), the parameters $m$ and $\lambda_{\phi}$
are computed as follows
\begin{align}
m = 0.014~\big(28542 + 264444 a - 30000 b - 6111 c - 10000\lambda\big)^{1/2}\label{m},
\end{align}
\begin{align}
\lambda_{\phi}=14.48~\big (0.95+8.814a -0.99b - 0.203c - 0.33\lambda\big)\label{lndphi}.
\end{align}
Therefore, utilizing the listed numerical ranges of parameters ($a$, $b$, $c$, and $\lambda$) in Table II,
the permitted amount of $m$ from Eq. (\ref{m}) is confined to place in the range of $2.42\leq m/M_{\rm p}\leq 2.67$.  Likewise, from Eq. (\ref{lndphi}) the parameter $\lambda_{\phi}$ is limited to be in the range of $13.00\leq\lambda_{\phi}\leq 15.71$.

\section{Conclusions}
\label{sec:V}
The approximate simultaneous observation of  GW170817 and GRB170817A events by LIGO-VIRGO collaboration, corroborate an infinitesimal  disparity around  ${\cal O}(10^{-16})$ between the promulgation velocity of GWs and the speed of light \cite{Abbott:2017d,Abbott:2017e}, thereunder that the value of $C_{T}$ is limited to be in the range of  $1-3\times10^{-15} \leq C_T\leq 1+7\times10^{-16}$ \cite{Abbott:2017e}. As a consequence, the gravitational theories which predict $C_{T}\neq 1$ for the propagated tensor perturbations must be reconsidered seriously.

We have inspected the viability of one of the modified gravity theories namely Higher-Order expansion of  Mimetic Gravity  model in view of GW170817-GRB170817A data. HOMimG model as a specific case of scalar-tensor theories exposes in the low energy limit of  Ho$\check{r}$ava-Lifshitz gravity (an appropriate candidate for quantum gravity) \cite{Cognola:2016gjy,Rinaldi:2016oqp,Diez-Tejedor:2018fue}. Instabilities of this model are eliminated through the obtained nonzero sound speed for scalar perturbation ($C_s \neq 0$) from definitely  fracturing the structure of the original Horndeski model. The action (\ref{cmsvz}) of HOMimG model consists of four parameters $a$, $b$, $c$, and $\lambda$.

Considering the propagating speed of scalar and tensor perturbations ($C_{s}$ and $C_{T}$) the restrictions of
$c>0$ and $a<1$ should be respected to liberate the theory from ghost problems. Furthermore the steadiness guarantee inflicts  theoretical  requirements as $0 \leq C_T^2 \leq 1$ and $0 \leq C_s^2 \leq 1$, in which the upper bound gives rise to certainty of subluminal dissemination of scalar and tensor perturbations \cite{Hirano:2017zox,Zheng:2017qfs,Cai:2017dyi,Takahashi:2017pje,Gorji:2017cai}.

With these aspects in mind, we tried to specify the permitted ranges for the  parameters of the HOMimG model ($a$, $b$,  and $c$) utilizing the observational constraint from GW170817-GRB170817A on $C_{T}$ (\ref{LIGO-VIRGO}), beside two mentioned theoretical constraints on $C_{T}^{2}$ and $C_{s}^{2}$ due to  the stability of the model (see the top box of Table \ref{t1} for numerical ranges of each parameter). In the following, we illustrated the parametric space of theses three parameters in Fig. \ref{fig1}.  Thereafter, so as to enhance the exactness of the parametric space we appointed the supplementary limitation of $\gamma$  (related to the latest observational value for the age of the universe) (\ref{gammarange}) on parameters (see Fig. \ref{fig2} for schemed parametric space and the bottom box of Table \ref{t1} for numerical result). In pursuance of determining the other parameter of the model ($\lambda$), another restriction on parameters is required and  the form of potential for the  model must be  specified. Hence,  we have selected  quadratic ($\frac{1}{2}m^{2}\phi^{2}$) and quartic ($\frac{1}{4} \lambda_{\phi} \phi^{4}$) potentials beside another observational restriction on the Equation of State parameter of dark energy (\ref{omega_df})  in addition to the erstwhile ones ( see Table \ref{t2} for numerical ranges of each parameters related to the selected potentials). In consequence we derived  practicable HOMimG model confined to numbers of observational and theoretical constraints. Moreover,  regarding the deduced numerical ranges for the model parameters summarized in Table \ref{t2}, and  cogitating two different potential  (quadratic and quartic potentials) to specify $\lambda$ parameter, we showed that this model is  independent of the form of potential. At last, utilizing the present value of density parameter of dark fluid (\ref{frac_density1}) and numerical ranges of model parameters listed in Table II, we could compute the permitted range of  $(2.42,2.67)$ for the parameter $m$, and  $(13.00,15.71)$ for $\lambda_{\phi}$  associated with quadratic and quartic potentials, respectively.

\section{Acknowledgements}\label{Ack}
The authors thank the anonymous referee for their valuable comments.


\end{document}